\documentclass[aps,prb,reprint,superscriptaddress,longbibliography]{revtex4-2}
\usepackage{graphicx} % Required for inserting images
\usepackage{amsmath,bbm}
\usepackage{hyperref}

\usepackage{tikz}
\usetikzlibrary{arrows.meta,decorations.pathreplacing}

\newcommand{\e}{\mathrm{e}}
\newcommand{\im}{\mathrm{i}}

\newcommand{\js}[1]{{\color{black}#1}}

\begin{document}

\title{Pseudospectral phenomena and the origin of the non-Hermitian skin effect}
\author{Jesko Sirker}
\affiliation{Department of Physics and Astronomy and Manitoba
Quantum Institute, University of Manitoba, Winnipeg, Canada R3T 2N2}
\date{\today}

\begin{abstract}
The non-Hermitian skin effect (NHSE), characterized by a macroscopic accumulation of eigenstates at the edge of a system with open boundaries, is often ascribed to a non-trivial point-gap topology of the Bloch Hamiltonian. We revisit this connection and \js{separate the question of the NHSE as a spectral reconstruction effect in clean systems from the question of stable topological protection.} For a Hatano-Nelson ladder, where point-gap winding and non-normality can be varied independently, we demonstrate that, \js{in a clean translationally invariant multiband setting}, the NHSE can occur without point-gap winding and, conversely, that point-gap winding can persist without the NHSE. These results establish that \js{in the clean case the connection between point-gap winding and the NHSE only holds in the scalar one-band case but, in general, not in the multiband case. Even more importantly,} the eigenspectrum of non-normal operators is \js{generically} highly sensitive to boundary conditions and perturbations, and therefore does not constitute a stable object encoding topological information. Instead, topological properties are reflected in the \js{splitting of the} singular-value spectrum for finite systems and, in the semi-infinite limit, correspond to boundary-localized \js{kernel modes} implied by the index of the corresponding Toeplitz operator.  
\end{abstract}

\maketitle

\section{Introduction}
Non-Hermitian systems exhibit a number of phenomena that have no counterpart in Hermitian quantum mechanics. Among these, the non-Hermitian skin effect, characterized by the macroscopic accumulation of eigenstates at the edge of a system with open boundaries, has attracted particular attention. The NHSE is often ascribed to a non-trivial point-gap topology of the corresponding Bloch Hamiltonian, where a nonzero spectral winding under periodic boundary conditions is taken as an indicator of boundary localization \cite{YaoWang,YokomizoMurakami,BergholtzBudich,Gong2018PRX,OkumaKawabata,KunstEdvardsson,ZhangYang}. \js{It is important to note though that this framework combines two physically distinct questions: first, is point-gap winding equivalent to the occurrence of the NHSE when a clean translationally invariant system is cut open and, second, what is the bulk-boundary correspondence which is stable to small generic perturbations and therefore genuinely topological?}

\js{The connection between point-gap topology and the NHSE} is largely motivated by simple \js{one-band} models such as the Hatano--Nelson chain \cite{HatanoNelson}, in which non-reciprocal hopping simultaneously gives rise to spectral winding and boundary-localized eigenstates. Generalized Brillouin zone constructions further reinforce this connection by providing a framework in which the open-boundary eigenspectrum can be reconstructed from a complexified Bloch Hamiltonian \cite{LeeThomale,YokomizoMurakami,Kawabata2019PRX}. \js{For the first question of boundary reconstruction in clean systems, this approach provides a natural framework. For the second question about stable topological objects, it relies, however, crucially on translational invariance.} This is conceptually problematic, since translational invariance is not a topological property. To the contrary, topological characteristics must be robust under perturbations that break translational symmetry \cite{RyuSchnyder}. A property that is equally important in non-Hermitian physics but which has received considerably less attention is that physically relevant non-Hermitian systems are typically described by non-normal operators whose eigenspectra are generically unstable under perturbations \cite{TrefethenEmbree2005,Davies2007,BoettcherSilbermann,OkumaSato,OkumaSato2,NakaiOkuma,BrunelliWanjura}.

This raises \js{the second, more} fundamental question: does the eigenspectrum of a non-normal operator provide a stable object that can encode topological information? In this work, we revisit this question by disentangling non-normality and spectral instability from point-gap topology using the Hatano-Nelson chain and ladder as examples. We explicitly show that the NHSE is a manifestation of the extreme sensitivity of non-normal operators to boundary conditions and perturbations, rather than a topological property. The topological content of the system is instead encoded in the index of the corresponding Toeplitz operator, which is associated with a true boundary-localized eigenstate of the semi-infinite system. For a finite system, the point-gap topology is reflected in the singular-value spectrum---which remains stable under perturbations---rather than in the eigenspectrum \cite{BoettcherSilbermann,HerviouBardarson,MardaniPimenta,Sirker_2dnh}. The role of the singular-value spectrum in non-Hermitian systems has also been emphasized in driven-dissipative systems \cite{BrunelliWanjura,WanjuraNunnenkamp2025}.

We show that intuition built on the Hatano-Nelson chain can be misleading because the non-reciprocity ratio controls both the point-gap topology and the non-normality of the model. Furthermore, the system with open boundaries is quasi-Hermitian, leading to a non-generic stability of its eigenspectrum with respect to certain perturbations that respect this quasi-Hermitian property. To disentangle spectral winding from non-normality, we therefore introduce a Hatano-Nelson ladder where these two aspects can be varied independently. This construction allows us to identify regimes with and without the NHSE in the presence or absence of point-gap winding. In particular, we demonstrate explicitly that the NHSE can occur without point-gap winding and, conversely, that point-gap winding can persist in the absence of the NHSE. These results establish that, \js{already within clean translationally invariant multiband systems,} the two phenomena are, in general, independent.

To summarize the main points we will make in this paper: (i) While translational invariance provides a useful tool for calculating topological indices \js{and the boundary reconstruction in clean one-band models}, the corresponding topological objects must be robust under its breaking. We show that this is not the case for the eigenspectrum of non-normal operators, whose strong sensitivity to perturbations underlies the NHSE. (ii) The topological content is instead encoded in the index of the associated Toeplitz operator, which determines the existence of boundary-localized modes in the semi-infinite system and, for finite systems, is reflected in the singular-value spectrum. This establishes that the NHSE should be understood as a specific manifestation of a more general spectral instability, which is distinct from the topological structure of the system.

Our paper is organized as follows: In Sec.~\ref{Sec_HN}, we present the general framework to analyze the NHSE and topology based on fundamental theorems, and exemplify those general results using the one-dimensional Hatano-Nelson model as an instructive example. We clarify the roles of non-normality, non-reciprocity, and point-gap topology, \js{and also briefly discuss the experimental implications}. In Sec.~\ref{Sec_ladder}, we introduce a Hatano-Nelson ladder and present explicit counterexamples showing that the NHSE \js{even in clean multiband systems} is in general not tied to point-gap topology. We conclude in Sec.~\ref{Sec_Concl} with a summary and discussion of our results.

\section{One-dimensional Hatano-Nelson model}
\label{Sec_HN}
Our results on the spectral instability of non-normal operators and the relation between topological indices, the kernels of Toeplitz operators, and the singular-value spectra of finite systems are fully general and based on well-established mathematical theorems. We discuss them here using a concrete example to show how they are applied in practice and to contrast them directly with the prevalent framework in the physics literature.

\subsection{General framework and rigorous results}
A lot of the intuition about non-Hermitian physics has been built on one of the simplest models, the Hatano-Nelson chain \cite{HatanoNelson}
\begin{equation}
    \label{HN}
    H=\sum_j \left[t_Rc_{j+1}^\dagger c_j + t_L c_j^\dagger c_{j+1} +\mu\, c_j^\dagger c_j \right] \, .
\end{equation}
We assume without loss of generality that $t_{R,L}>0$. For open boundary conditions (OBC) the Hamiltonian has a Toeplitz form in real space
\begin{equation}
\label{Toplitz}
    \mathcal{H}= \begin{pmatrix}
    h_0 & h_1 & h_2 & \cdots \\
    h_{-1} & h_0 & h_1 & \cdots \\
    h_{-2} & h_{-1} & h_0 & \cdots \\
    \vdots & \vdots & \vdots & \ddots
    \end{pmatrix}
\end{equation}
with $h_0=\mu$, $h_1=t_L$ and $h_{-1}=t_R$, and all other matrix elements equal to zero. If we define the Bloch Hamiltonian---which is also called the symbol in the mathematical literature---as 
\begin{equation}
    \label{1dsymbol}
    h_j=\frac{1}{2\pi}\int_0^{2\pi}dk\, h(k)\e^{ikj} \,
\end{equation}  
then we obtain
\begin{eqnarray}
    \label{hk}
    h(k)&=&\mu+t_R\e^{ik}+t_L\e^{-ik} \nonumber \\
    &=& \mu +(t_R+t_L)\cos k +\im(t_R-t_L)\sin k \, .
\end{eqnarray}
For non-reciprocal couplings, $t_R\neq t_L$, $h(k)$ describes an ellipse in the complex plane centered at $\mu$ with semi-axes $t_R+t_L$ and $|t_R-t_L|$. If an energy $E$ is inside the ellipse, then $h(k)$ exhibits a winding with respect to this reference energy which is captured by the winding number
\begin{equation}
    \label{winding}
    \mathcal{I}(E)=\frac{1}{2\pi \im}\int_0^{2\pi}dk\, \partial_k \ln(h(k)-E)\, .
\end{equation}
For $E$ inside the ellipse and $t_R>t_L$ we have $\mathcal{I}(E)=+1$, while $\mathcal{I}(E)=-1$ if $t_L>t_R$. The winding vanishes if $E$ is outside the ellipse. 

One of the most striking features of the model is that for periodic boundary conditions (PBC) the eigenvalues lie on the ellipse in the complex plane described by $h(k)$ with $k_n=2\pi n/N$, where $N$ is the system size and $n=0,\cdots,N-1$, while the corresponding eigenstates are all bulk states. This changes dramatically when open boundary conditions (OBC) are imposed: In this case, the eigenvalues are all real and the corresponding eigenstates are all localized either at the right or left boundary depending on which of the hopping amplitudes dominates. This high sensitivity of the spectrum with respect to the boundary conditions is very different from what we are used to in Hermitian systems and is called the non-Hermitian skin effect (NHSE). Because the NHSE occurs for non-reciprocal couplings $t_R\neq t_L$ when the Bloch Hamiltonian always winds, the dominant framework for understanding the NHSE is that it is of topological origin. Most prominently, generalized Brillouin zone constructions \cite{YaoWang,YokomizoMurakami} reproduce the open-boundary eigenspectrum by analytically continuing the Bloch momentum $k\to k+i\kappa$. The central question, however, is not whether the open-boundary eigenspectrum \js{of a clean system} can be reconstructed from a modified Bloch Hamiltonian, but whether the eigenspectrum itself is the appropriate spectral object to encode \js{stable} topological information.

At this point it is therefore important to pause and to carefully distinguish different aspects of non-Hermitian systems that are, in general, independent but appear connected in the simple Hatano-Nelson model. The three relevant aspects are
\begin{itemize}
    \item Non-normality, $H^\dagger H \neq H H^\dagger$,
    \item Non-reciprocity, $t_R\neq t_L$,
    \item and topological winding, $\mathcal{I}(E)\neq 0$.
\end{itemize}
As we will show below, these three properties play conceptually different roles: non-reciprocity and non-normality determine the spectral sensitivity and are responsible for the NHSE, while the winding number determines the index of the associated Toeplitz operator and the existence of topologically protected boundary modes for the semi-infinite chain. The point least discussed in the physics literature is that physically interesting non-Hermitian matrices are also most often non-normal. This includes, in particular, models with non-reciprocal couplings. It is this non-normality, and not the non-Hermiticity per se, which fundamentally alters the spectral properties. While for a normal matrix---which includes the Hermitian systems we are most used to in quantum physics---the Bauer-Fike theorem \cite{TrefethenEmbree2005,HornJohnson2013} guarantees the stability of the eigenspectrum, the spectrum of a non-normal matrix is highly sensitive to small changes such as a change in boundary conditions or a breaking of translational invariance. It is also important to remember in this context that topological protection is about the stability of parts of the spectrum with respect to perturbations. The Bloch Hamiltonian is merely a convenient tool to calculate such properties but topological properties have to persist even if translational invariance is broken \cite{RyuSchnyder}. The common identification of spectral winding with topological properties of the eigenspectrum therefore implicitly relies on translational invariance. This is a stronger assumption than topological protection and does not hold in generic settings. As a consequence, the eigenspectrum of non-normal operators cannot, in general, be used to infer topological properties.

The topological properties of the system are, in general, independent from the spectral instability due to the non-normality of the operator and have their foundations in Toeplitz operator theory. If a system has a winding \eqref{winding} with respect to some reference energy $E$, then this defines a topological index via $\mbox{ind}(H-E)=-\mathcal{I}(E)$. What this index relates to fundamentally is the dimension of kernels, $\mbox{ker}(H)=\{v|Hv=0\}$ of semi-infinite chains, more precisely
\begin{equation}
    \label{index}
    \mbox{ind}(\tilde H) = \mbox{dim}(\mbox{ker}(\tilde H)) - \mbox{dim}(\mbox{ker}(\tilde H^\dagger)) \, ,
\end{equation}
where we have defined $\tilde H=H-E$. It is important to note that the index is always zero for finite systems (rank-nullity theorem), and thus is a genuine property of the semi-infinite system. For a finite system, the topological index, in general, does not connect to properties of the unstable eigenspectrum but rather to the properties of the stable singular value spectrum \cite{BoettcherSilbermann}. The latter is given by $s_i(\tilde H)=\sqrt{\lambda_i(\tilde H^\dagger \tilde H)}$, i.e., the singular values are the square roots of the eigenvalues of $\tilde H^\dagger \tilde H$ which is a Hermitian matrix with a stable eigenspectrum. If $\tilde H$ itself is Hermitian, then the singular values are just the absolute values of the eigenvalues of $\tilde H$. In this case, one can synonymously talk about the topological properties of the singular-value spectrum or the eigenspectrum. Since we are used to dealing with Hermitian operators in quantum physics, this has sometimes led to the incorrect assumption that it is always the eigenspectrum which shows topological properties if a non-zero topological index exists. The proper general connection between the topological index and the singular-value properties of a finite system is given by $K$-splitting theorems \cite{BoettcherSilbermann} which state that for a system with winding $\mathcal{I}(E)=-\mbox{ind}(\tilde H)$, the Hamiltonian $\tilde H$ has at least $K\geq |\mathcal{I}(E)|$ singular values that are separated from the bulk spectrum by a gap and go to zero in the thermodynamic limit. There is also one other aspect that is different when considering Hermitian versus non-normal systems: In a Hermitian system, we are used to having finite truncations of topologically protected boundary modes which exponentially converge to a zero mode in the semi-infinite limit, $\tilde H|\Psi\rangle\sim\e^{-N}|\Psi\rangle$. Here $N$ is the system size. In a non-normal system, this is in general replaced by $\| \tilde H|\Psi\rangle\|\sim \e^{-N}$ but $|\Psi\rangle$ will typically only be an exact eigenstate in the semi-infinite limit. I.e., the topologically protected modes are not visible when considering the eigenspectrum of a finite system. Nevertheless, these topologically protected boundary modes remain physically highly relevant. They correspond to metastable states but with lifetimes which make them practically indistinguishable from true eigenmodes in macroscopic systems, see also the Suppl.~Mat.~of Ref.~\cite{MonkmanSirker_nH}. \js{We discuss the experimental implications further at the end of this section.}

In the Hatano-Nelson model, the in-principle independent aspects of non-normal spectral instability and topological winding are directly intertwined by the non-reciprocity of the couplings. It is the non-reciprocity that makes the Hamiltonian non-normal while, at the same time, also making $h(k)$ wind. The reason is that this simple one-band model has only one relevant tuning parameter, the ratio $t_R/t_L$. Further complicating the attempt of understanding non-Hermitian systems based on the intuition gained from the Hatano-Nelson model is that this model with OBC is similar to a symmetric (real Hermitian) operator. We can diagonalize the open Hatano-Nelson model by first performing a similarity transform $H'=D^{-1}HD$ with $D=\mbox{diag}(1,\sqrt{t_R/t_L},t_R/t_L,\cdots)$\js{---which is precisely the imaginary gauge transformation underlying the GBZ deformation $k\to k+i\kappa$ with
$\kappa=\frac{1}{2}\ln(t_R/t_L)$, for which $h(k+i\kappa)=2\sqrt{t_Rt_L}\cos k$ reproduces the OBC spectrum---}followed by a diagonalization $H^d=U^T H'U$ with
\begin{equation}
    \label{OBC1}
   U_{ij}=\sqrt{\frac{2}{N+1}}\sin\left(\frac{\pi}{N+1}ij\right) \, .
\end{equation}
\js{Note that the similarity transformation $D$ is not unitary and that the general instability of the OBC eigenspectrum is reflected in the condition number $\sim [\max(t_R,t_L)/\min(t_R,t_L)]^{N/2}$ diverging exponentially for $N\to\infty$.} The eigenvalues are then given by
\begin{equation}
    \label{OBC2}
    \lambda_n = \mu+2\sqrt{t_R t_L}\cos\left(\frac{\pi}{N+1}n\right)
\end{equation}
with $n=1,\cdots,N$. As a consequence, the eigenvalues are real and, \js{while unstable to generic perturbations, do remain real and relatively stable under local perturbations respecting the quasi-Hermiticity}. Because of these quasi-Hermitian properties, the Hatano-Nelson chain provides a sometimes misleading intuition about the spectral behavior of generic non-normal matrices. 

The general tool to investigate the spectral instability of non-normal matrices is the pseudospectrum, which is defined as
\cite{TrefethenEmbree2005,Davies2007}
\begin{equation}
    \label{pseudo}
    \sigma_\varepsilon(H) = \{ \lambda : \| (H-\lambda\mathbf{1})^{-1} \|_2 > 1/\varepsilon \} \, ,
    \end{equation}
where $\|\cdots\|_2$ is the operator norm. For normal matrices, the pseudospectrum is given by $\sigma_\varepsilon(H) =\{\lambda: \mbox{dist}(\lambda,\sigma(H)) < \varepsilon \}$, i.e., the eigenspectrum $\sigma(H)$ is merely thickened by all values within a range $\varepsilon$ of an eigenvalue. For non-normal matrices $H$, in contrast, the pseudospectrum contains points that can be very far apart from any true eigenvalue. For a scalar Bloch Hamiltonian $h(k)$, there are rigorous theorems \cite{BoettcherSilbermann} showing that in the thermodynamic limit, the pseudospectrum is controlled by the image of $h(k)$ and, for the one-dimensional Hatano-Nelson model, is contained \js{within an $\varepsilon$-neighborhood of} the region enclosed by the ellipse traced by $h(k)$, converging to this region
for $\varepsilon\to 0$. In practice, however, generic perturbations do not populate this region uniformly. Instead, most eigenvalues are typically found in a thickened band near the curve $h(k)$. We note that two-dimensional scalar Bloch Hamiltonians $h(k_x,k_y)$ depend on two momenta and thus the pseudospectrum is inherently two-dimensional and occupies a finite area in the complex plane, see, for example, Fig.~2(c) in Ref.~\cite{Sirker_2dnh} where the pseudospectrum of the two-dimensional Hatano-Nelson model is shown. 

\subsection{Numerical results for the eigensystem}
\begin{figure}
    \centering
    \includegraphics[width=0.99\columnwidth]{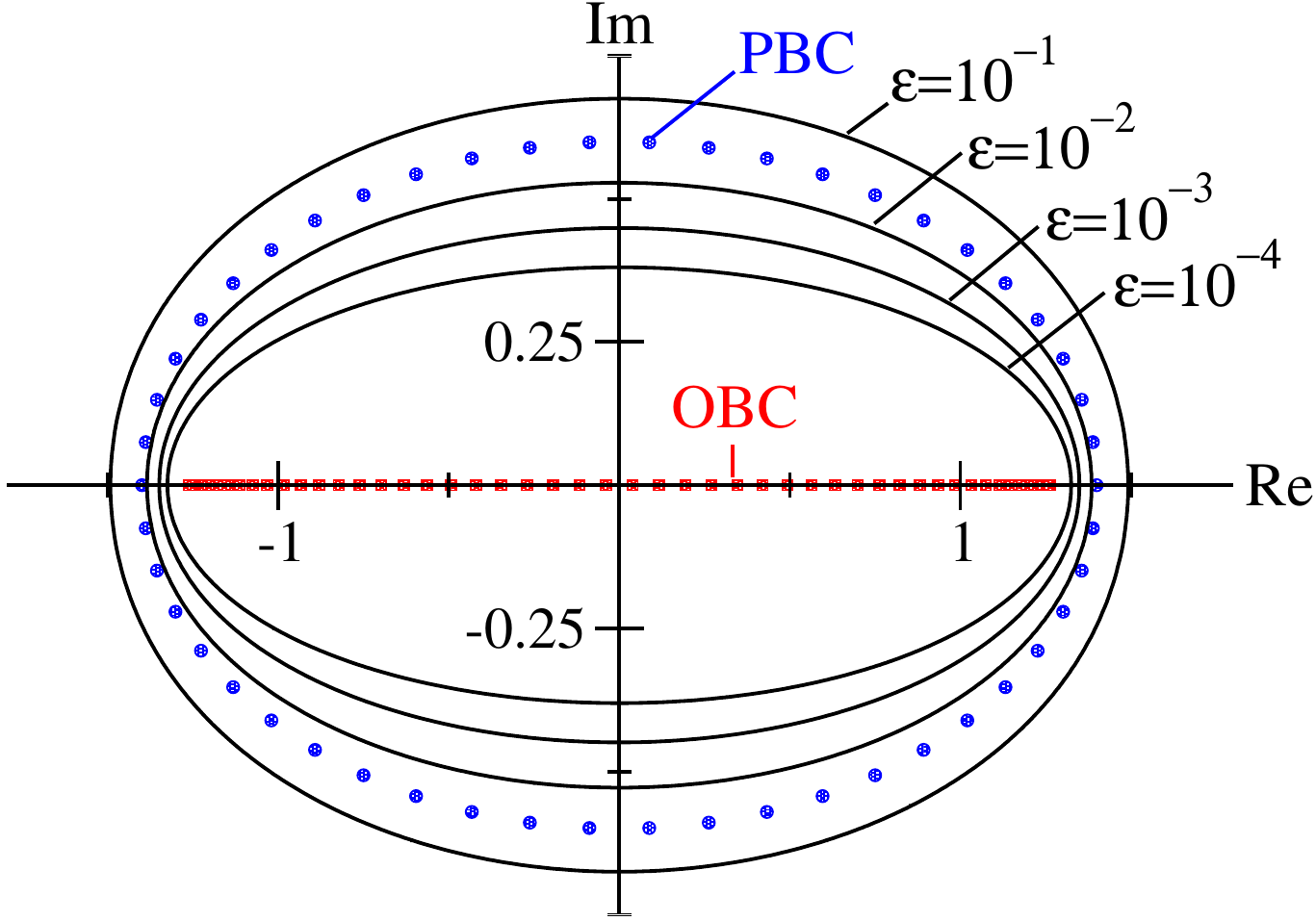}
    \caption{Spectrum of the Hatano-Nelson model with $N=50$ and $t_L=1$, $t_R=0.4$, $\mu=0$ for PBC and OBC. Also shown are the pseudospectral contour lines $\|(H-\lambda\mathbf{1})^{-1}\|_2=1/\varepsilon$ for the open chain, which show the extreme instability of the eigenspectrum.}
    \label{Fig1}
\end{figure}

To support these rigorous results, we show in Fig.~\ref{Fig1} numerical data for a Hatano-Nelson chain with $N=50$ sites.
For PBC, the spectrum is given by $h(k)$ with $k=2\pi n/N$ and $n=0,\cdots,N-1$. For OBC, on the other hand, the spectrum is purely real and given by Eq.~\eqref{OBC2}. The figure also shows the pseudospectrum of the open chain, Eq.~\eqref{pseudo}, for different $\varepsilon$. The results demonstrate that the spectrum is extremely sensitive to perturbations. The pseudospectrum extends into the complex plane, and for small $\varepsilon$ its contours lie close to $h(k)$. I.e., the spectrum of the open chain under generic small perturbations tends to accumulate near the Bloch curve in the thermodynamic limit. Numerically, this is also immediately obvious if one tries to diagonalize the Hamiltonian matrix of large open chains in standard double-precision arithmetic. What one obtains then is not the true spectrum but rather a sampling of the pseudospectrum with eigenvalues accumulating near $h(k)$ because of the double precision errors essentially acting as tiny random perturbations. Based on the spectral instability of the open-chain spectrum, it should be clear that there cannot be any topological protection present in the eigenspectrum. Topological features cannot be related to translational invariance, and have to be stable under generic perturbations \cite{RyuSchnyder}.

\begin{figure}
    \centering
    \includegraphics[width=0.99\columnwidth]{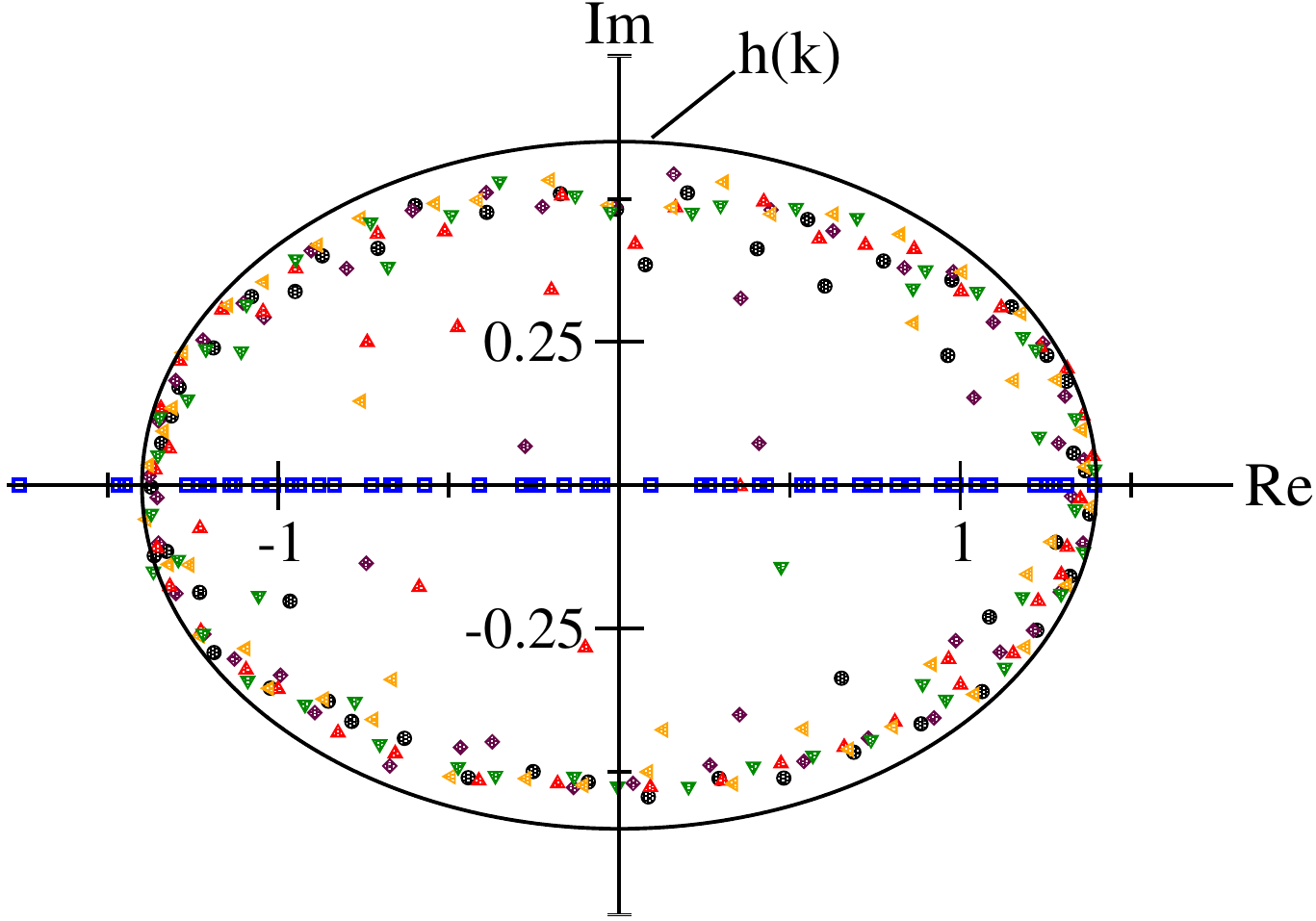}
    \caption{Perturbed spectrum of the open Hatano-Nelson model for $N=50$ and $t_L=1$, $t_R=0.4$, $\mu=0$. The open blue squares on the real axis are obtained by perturbing each individual onsite potential and hopping by a \js{real} additive random perturbation drawn from the interval $[-0.4,0.4]$. The other symbols denote five realizations of the Hamiltonian perturbed by a random complex matrix $A$ with $|a_{ij}|\leq 0.1/\sqrt{N}$.}
    \label{Fig2}
\end{figure}

This fact is somewhat obscured in the Hatano-Nelson model with OBC because its Hamiltonian is then quasi-Hermitian so that \js{certain} local perturbations, which keep the quasi-Hermiticity of the Hamiltonian intact, probe this instability only weakly. This is demonstrated in Fig.~\ref{Fig2}, where we compare generic random perturbations with \js{real} perturbations that only affect the onsite potential $\mu$ and the hopping amplitudes $t_{R/L}$.
\js{We note that complex perturbations of $\mu$ and $t_{R/L}$ break the reality of the spectrum, which moves into the complex plane by $\mathcal{O}(\varepsilon)$.}  \js{Furthermore}, for a generic perturbation---implemented by adding a random complex matrix to the Hamiltonian---the \js{full} spectral instability is already obvious for the considered modest system size and weak perturbation strength, with the spectrum starting to accumulate near the $h(k)$ ellipse. Note that for a dense random perturbation, maintaining a perturbation of $\mathcal{O}(1)$ operator norm requires matrix elements of $\mathcal{O}(1/\sqrt{N})$. 

\js{We note that a small dense perturbation of the Hamiltonian matrix is physically realistic because a model with short-range interactions will always only be an approximation of an experimental realization of a non-Hermitian system. No matrix element is exactly zero in a real physical system unless enforced by
symmetries or exact conservation laws. In particular, background perturbations of constant magnitude including small longer-range couplings are unavoidable. Such constant perturbations, no matter how small, will eventually be large compared to the $\mathcal{O}(1/\sqrt{N})$ matrix elements used in our dense perturbation ensemble, and will therefore act as a generic perturbation for sufficiently large systems.} We conclude that the pseudospectrum characterizes the existence of experimentally-relevant destabilizing perturbations, not the behavior of every perturbation ensemble. For the open Hatano–Nelson chain, local random perturbations \js{which do not violate its quasi-Hermitian structure} probe this instability only weakly, even though the resolvent norm is already large.

\begin{figure}
    \centering
    \includegraphics[width=0.99\columnwidth]{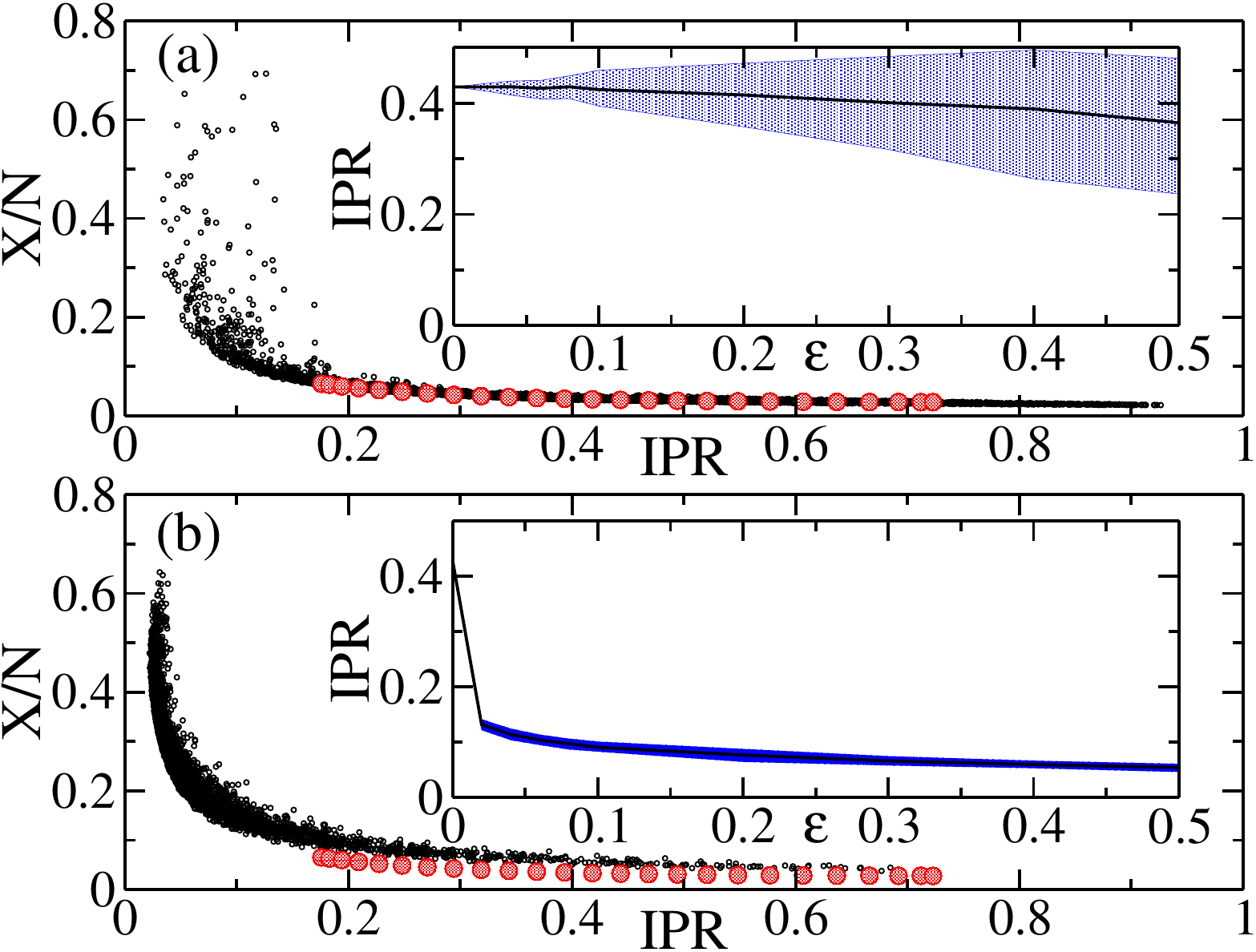}
    \caption{Hatano-Nelson model with OBC and parameters as in Fig.~\ref{Fig2}. The main panels show the center $X/N$ of each wave function versus its IPR, providing a simultaneous measure of localization strength and spatial bias. The large red dots represent the unperturbed system and the black circles in (a) $100$ realizations of the \js{real tridiagonal} perturbation with strength $\varepsilon=0.2$, and in (b) $100$ realizations of the global perturbation with strength $0.2/\sqrt{N}$. The insets show the mean IPR as a function of disorder strength $\varepsilon$ with the blue shaded bands showing the $5-95$ percentile range.}
    \label{Fig3}
\end{figure}

This difference becomes even more obvious when considering the spatial structure of the eigenstates, shown in Fig.~\ref{Fig3}. For each eigenstate $\Psi$, we calculate the inverse participation ratio (IPR), which is defined as $I=\sum_{n} |\Psi(n)|^4$ as well as the center of the wave function $X=\sum_n n |\Psi(n)|^2$ \js{where we use the normalization $\sum_n |\Psi(n)|^2=1$}. For a completely delocalized mode, we will have $I\sim 1/N$ while $I\sim\mathcal{O}(1)$ for a localized mode. The center of the mode $X$ will then tell us where most of the spectral weight of this mode is located. For the unperturbed system, we find that all eigenstates have an IPR $I\sim \mathcal{O}(1)$ and that their centers are all located at the left boundary consistent with the NHSE. For the perturbed system, we calculate $I$ and $X$ for all eigenstates in a disorder realization and do so for $100$ realizations. For the \js{real tridiagonal} perturbation, shown in Fig.~\ref{Fig3}(a), most modes remain localized at the left boundary but a fraction of states become significantly more extended with a center no longer located at the left boundary. This is reflected in the IPR averaged over all states shown in the inset, which only decays very slowly with increasing disorder strength but has a band of possible IPRs (shown is the $5$ to $95$ percentile range) that grows significantly with perturbation strength $\varepsilon$. In contrast, even weak global perturbations delocalize most of the states as shown in Fig.~\ref{Fig3}(b). The average IPR almost immediately collapses to values close to $1/N$ and the centers of most of the wave functions are no longer situated at the left edge, demonstrating almost complete delocalization. 
We conclude that the relative stability of the spectrum to \js{certain} local perturbations is due to the quasi-Hermiticity of the Hamiltonian---a property that still holds for the perturbed system if the matrix remains 
real tridiagonal \js{and diagonally symmetrizable}---and is visible both in the eigenspectrum, which remains real, as well as in the eigenstates, which remain largely localized. We stress again that this is a peculiar property of the rather simple single-band Hatano-Nelson model with nearest-neighbor hopping and not indicative of the general behavior of non-normal Hamiltonians. 

\subsection{Topological properties}
\label{II_top}
Clearly, the skin modes investigated in the previous section are unstable under generic perturbations and thus not topologically protected. However, for $t_L>t_R$ and $\mu=0$ as chosen in the numerically studied examples, the winding number is given by $\mathcal{I}(E=0)=-1$, and we do have a non-zero topological index. What is topologically protected, then?

First, if we have a non-zero winding $\mathcal{I}(E)=\pm 1$ (other windings are not possible in this specific model), then this implies that the index defined in Eq.~\eqref{index} is non-zero. This means that there are vectors that lie in the kernel of the semi-infinite $\tilde H=H-E$ or of the adjoint operator. For a scalar Bloch Hamiltonian, as in the Hatano-Nelson model, Coburn's lemma \cite{BoettcherSilbermann} furthermore dictates that the kernels of $\tilde H$ and $\tilde H^\dagger$ cannot be non-zero at the same time. For non-zero winding $\mathcal{I}(E)=\pm 1$, the semi-infinite Hatano-Nelson model described by $\tilde H$ therefore has either a right or a left exact zero-energy boundary localized mode. This is the fundamental connection between the topological index and the existence of boundary modes.

Second, because of the simplicity of the Hatano-Nelson model, we can demonstrate the connection between the winding $\mathcal{I}(E)$ and the kernel modes of $\tilde H$ and $\tilde H^\dagger$ explicitly. If we define $\tilde\mu=\mu-E$, then we have to solve $\tilde H|\Psi\rangle =0$ with $\tilde H$ as given in Eq.~\eqref{Toplitz} and $h_0=\tilde\mu$, $h_1=t_L$, and $h_{-1}=t_R$. This leads to the following bulk and boundary equations
\begin{eqnarray}
    \label{recur1}
    && \tilde\mu \Psi_1 + t_L \Psi_2 = 0 \\
    && t_R \Psi_j+ \tilde\mu\Psi_{j+1} + t_L\Psi_{j+2}=0,\; j\geq 1 \nonumber \, . 
\end{eqnarray}
If we make the ansatz $\Psi_j=\lambda^j$, then the bulk equation is a quadratic equation, $p(\lambda)=t_L\lambda+\tilde\mu + t_R\lambda^{-1}=0$, which has two solutions, $\lambda_\pm$. A general solution is thus of the form
\begin{equation}
    \label{recur2}
    \Psi_j = A\lambda_-^j + B\lambda_+^j
\end{equation}
where the coefficients $A,B$ are determined by the boundary condition in Eq.~\eqref{recur1} plus the normalizability condition $\sum_j |\Psi_j|^2=1$. The latter can only be satisfied if both solutions of the quadratic equation $p(\lambda)$ are inside the unit disk, $|\lambda_\pm|<1$. Crucially, it is the winding number $\mathcal{I}$ that tells us where the two solutions are located in the complex plane. Using Cauchy's argument principle we have
\begin{eqnarray}
    \label{recur3}
    N_{\textrm{in}}-N_p &=&\frac{1}{2\pi\im}\oint_{|z|=1}\frac{p'(z)}{p(z)}dz \\
    &=& -\frac{1}{2\pi\im}\int_0^{2\pi} dk\, \partial_k\ln(h(k)-E)=-\mathcal{I}(E) \nonumber
\end{eqnarray}
where $N_{\textrm{in}}$ is the number of zeroes inside the unit circle and $N_p$ the number of poles. We have also used the definition of the winding number \eqref{winding} and that $p(z)=h(z)$ with $h(z)$ being the Bloch Hamiltonian as function of $z=\e^{-\im k}$. Note that $p(z)$ has a single pole inside the disk, $N_p=1$, which leads to
\begin{equation}
    \label{recur4}
    N_{\textrm{in}} = 1-\mathcal{I}(E) \, .
\end{equation}
We therefore conclude that both solutions $\lambda_\pm$ are inside the unit disk and a normalizable boundary localized solution of $\tilde H|\Psi\rangle=0$ exists if and only if $\mathcal{I}(E)=-1$. If, on the other hand, $\mathcal{I}(E)=0$ then one solution is inside and one outside the unit disk and thus no normalizable solution exists. Finally, if $\mathcal{I}(E)=+1$ then both solutions are outside the unit disk and therefore a normalizable boundary localized solution of the adjoint operator, $\tilde H^\dagger|\Psi\rangle=0$, exists. This explicitly demonstrates the index theorem \eqref{index} in this case.

The explicit solution of the quadratic equation \eqref{recur1} is given by
\begin{equation}
    \label{recur5}
    \lambda_\pm =\frac{-\tilde\mu\pm\sqrt{\tilde\mu^2-4 t_Lt_R}}{2t_L}
\end{equation}
and using the boundary condition in Eq.~\eqref{recur1}, we find the explicit solution for the boundary mode
\begin{equation}
    \label{recur6}
    \Psi_j = A(\lambda_+^j -\lambda_-^j)
\end{equation}
which is a normalizable solution if $\mathcal{I}(E)=-1$ ($t_L>t_R$). The normalized solution takes a particularly simple form and still shows all the physics if we consider the special case $\tilde\mu=0$. Then, the normalized solution is given by
\begin{equation}
    \label{recur7}
    \Psi_{2j}=0,\qquad \Psi_{2j+1}=\sqrt{1-(t_R/t_L)^2}\left(-\frac{t_R}{t_L}\right)^j \, .
\end{equation}
For a finite system with $N$ sites, the additional boundary condition at the other end of the chain will in general force the entire vector to be zero. The protected boundary mode is thus not an exact eigenvector in a generic finite system. Instead, the vector has to be cut off, which means it remains exponentially close to an eigenstate with $\|\tilde H|\Psi\rangle\|\sim (t_R/t_L)^{N/2}$. For a macroscopic number of sites, this state will thus be practically indistinguishable from a true zero-energy boundary-localized eigenstate {\it but it cannot be identified by the exact diagonalization of small systems}.

\begin{figure}
    \centering
    \includegraphics[width=0.99\columnwidth]{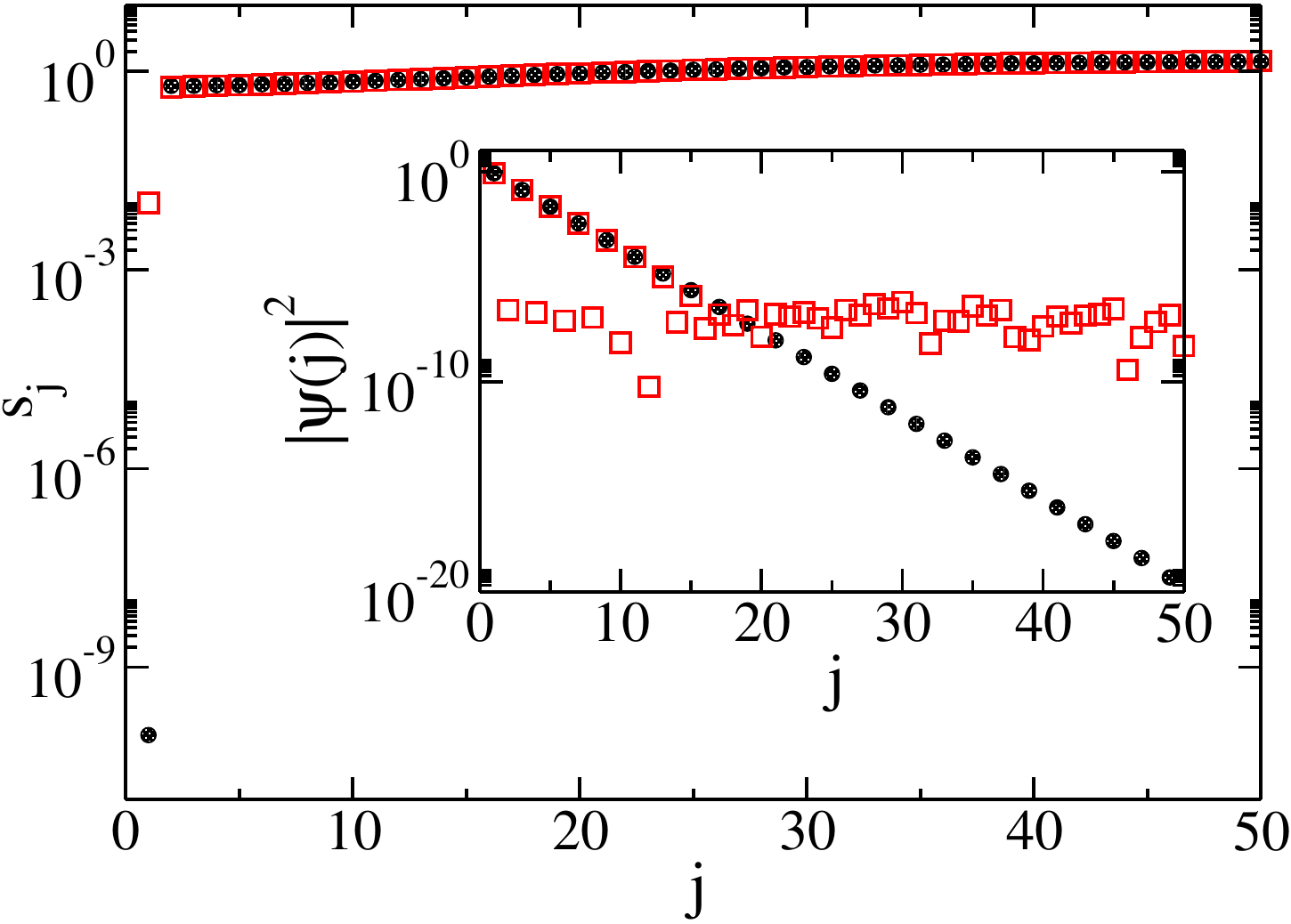}
    \caption{Singular value spectrum for the open Hatano-Nelson model with $N=50$, $t_L=1$, $t_R=0.4$, $\mu=0$. Circles denote the unperturbed system, open squares the system with a global perturbation of strength $0.1/\sqrt{N}$ with an average taken over $1000$ realizations. The topologically protected singular value near zero remains separated from the bulk spectrum by a gap. Inset: The corresponding right singular vector is exponentially localized at the left edge.}
    \label{Fig4}
\end{figure}

The third aspect is therefore how to identify these protected boundary modes if they are not visible in the eigenspectrum of a finite system. Here, so-called K-splitting theorems provide an answer which state that for systems with a scalar Bloch Hamiltonian, such as the Hatano-Nelson model, there are exactly $K=|\mathcal{I}|$ singular values which go to zero with increasing system size and which are separated from the bulk spectrum by a gap. The non-scalar case is slightly more complicated and discussed in detail in Refs.~\cite{MonkmanSirker_nH,Sirker_2dnh}. In Fig.~\ref{Fig4}, the singular value spectrum of the Hatano-Nelson model for the same parameters as in Fig.~\ref{Fig2} is shown.
Clearly visible is the topologically protected singular value, which is exponentially small in system size and separated from the bulk spectrum by a gap. This spectral gap is given in the thermodynamic limit by $\Delta_0=\min_k |h(k)|=|t_R-t_L|$. Any perturbation that is small compared to $\Delta_0$ will keep the splitting intact. The global perturbation shown in Fig.~\ref{Fig4} reduces the splitting but the topologically protected singular value remains well separated from the bulk. In the inset of Fig.~\ref{Fig4}, the right singular vector belonging to the topologically protected singular value is shown. This vector is localized at the left edge. For the unperturbed system, only the vector coefficients on odd sites are non-zero. The global perturbation averaged over $1000$ realizations induces a 'background' $|\Psi(j)|^2\sim 10^{-7}$ on both even and odd sites but the mode is still localized at the left edge and shows exponential decay up to the point where the background is reached.

The topologically protected singular vector $v$ in the unperturbed system is the solution of $\tilde H^\dagger \tilde H v= s^2 v$ with $s\to 0$ for $N\to\infty$. For $\tilde\mu=0$, the bulk recursion relation becomes 
\begin{equation}
    \label{recur8}
    t_R t_L v_{j-2} + (t_R^2 + t_L^2) v_j + t_R t_L v_{j+2} = 0 
\end{equation}
and the solutions on the even and odd lattice sites again separate. Taking the boundary conditions into account, one finds that $v_{2j}=0$ and $v_{2j+1}\sim (t_R/t_L)^j$, i.e., the vector is indeed exponentially localized at the left edge. The important point why this vector survives with small modifications even in a finite system is that $\tilde H^\dagger \tilde H$ is a Hermitian operator, so the eigensystem is stable, and small perturbations as well as changes in the boundary conditions do not drastically alter the eigenvectors in contrast to the non-Hermitian case. The singular-value spectrum of a finite system therefore contains all the information about the topology of the model whereas the eigenspectrum does not. \js{We note that the values $\pm s_i(\tilde H)$ coincide with the eigenspectrum of the doubled Hermitian Hamiltonian appearing in Refs.~\cite{OkumaKawabata,OkumaSato}; there, however, this construction is used only as an auxiliary tool to characterize eigenmodes of $\tilde H$, whereas the singular-value formulation makes explicit that the stable Hermitian spectral problem itself---and not the eigenspectrum of $\tilde H$---encodes the topology of the finite system.}

\subsection{Boundary reconstruction in clean systems}
\label{II_D}
\js{We have seen that the skin eigenmodes are not topologically protected and are destroyed by small generic perturbations. Nevertheless, one can ask the question if a clean system with non-zero point-gap winding always has skin modes when cut open in the same way as the Hatano-Nelson model. We stress that this is not a question about stable topological protection but rather a question about the reconstruction of eigenstates in clean systems when switching from PBC to OBC. To address this much more narrow question, the GBZ formalism is useful. We note that the equivalence of point-gap winding and skin modes has been proven in Ref.~\cite{ZhangYang} for the scalar one-band case. Here we present the main steps in a modified way to clarify why this proof does not generalize to the multiband case. This directly leads to the counterexamples discussed in Sec.~\ref{Sec_ladder}.

For a scalar one-band model with finite-range hopping, the Bloch Hamiltonian has the form 
\begin{equation}
    \label{hz}
    h(z)=\sum_{r=-p}^q t_r z^r 
\end{equation}
with $z = \e^{\im k}$, opposite to the convention $z=\e^{-\im k}$ of Eq.~\eqref{recur3}, which reverses the sign of the winding in the root count. For an energy $E$, we can write $h(z)-E = z^{-p}P_E(z)$ where $P_E(z)$ is a polynomial in $z$ of degree $p+q$. We denote the ordered roots of this polynomial by $|z_1(E)|\leq\cdots\leq |z_{p+q}(E)|$. The standard scalar GBZ construction starts from the bulk ansatz $\Psi_n=\sum_j c_j z_j^n$. Imposing the open-boundary conditions and taking the thermodynamic limit leads to the condition $f(E)\equiv|z_p(E)|=|z_{p+1}(E)|$, which determines the OBC spectral continuum. If $f(E)=1$ then the solution is extended, otherwise it is a mode localized either at the right or at the left boundary. By the argument principle, the winding around some energy $E_0$ is given by $\mathcal{I}(E_0)=N_\textrm{in}-p$ where $N_\textrm{in}$ is the number of roots of $P_{E_0}(z)$ inside the unit circle $S^1$. If $f(E)\equiv 1$ for all $E$ then the GBZ is $S^1$ and $N_\textrm{in}=p$ leading to $\mathcal{I}(E_0)=0$. Conversely, if the GBZ is not $S^1$ then one can find an energy $E_0$ with $\mathcal{I}(E_0)\neq 0$ and there is an exponentially localized solution \cite{ZhangYang}. In the scalar one-band case, if one can find an energy $E_0$ with non-zero point-gap winding then the following relations hold
\begin{equation}
    \label{scalar}
  \mathcal{I}(E_0)\neq 0 \Leftrightarrow\mbox{GBZ}\neq S^1\Leftrightarrow\mbox{macroscopic skin modes.}  
\end{equation}
However, this result does not generalize to multiband models. Here, the quantity determining the point-gap winding $\mathcal{I}(E)$ and thus the total root distribution is $\det[h(z)-E]$ which by itself no longer determines whether eigenstates are localized or extended. For example, eigenstates can now have an internal structure where parts are localized while other parts are extended. Therefore, in general, the existence of an energy $E_0$ with $\mathcal{I}(E_0)\neq 0$ no longer implies the existence of a macroscopic number of skin modes nor does the existence of skin modes imply the existence of an energy $E_0$ with $\mathcal{I}(E_0)\neq 0$. In Sec.~\ref{Sec_ladder}, we provide examples of a clean multiband system that demonstrate this point explicitly. Finally, we stress once more that even in the scalar one-band case, the theorem \eqref{scalar} is a statement about the reconstruction of the energy spectrum when cutting a clean system open. It does not imply that the skin modes are stable topological objects.
}

\subsection{Experimental implications}
\js{While Sec.~\ref{II_top} answers the question how to numerically identify topological boundary modes in a finite system, it leaves open what the experimental implications of a singular boundary vector are which belongs to a singular value that vanishes in the thermodynamic limit and which can exist in general without any eigenvalues approaching zero. Since the topological boundary mode $|v\rangle$ is exponentially close to an eigenmode, $\|\tilde H|v\rangle\|\sim \e^{-N}$, preparing a system in this state will result in a lifetime that increases with system size $N$. This could, for example, be tested in a Loschmidt echo experiment, see the Suppl.~Mat.~of Ref.~\cite{MonkmanSirker_nH}. 

Another experimental approach is to consider the linear response when driving a system with a topological boundary mode. Potential platforms for such an experiment are topoelectric circuits. Generally, we can consider a source profile $|s\rangle$ and the induced response $|\Psi\rangle$ which is given by
\begin{equation}
    \label{response}
    (H-E)|\Psi\rangle = |s\rangle\quad \Rightarrow \quad |\Psi\rangle = \underbrace{(H-E)^{-1}}_{G(E)}|s\rangle
\end{equation}
where $G(E)$ is the Green's function. If the Hamiltonian matrix is diagonalizable, we can use an eigen-decomposition and write 
\begin{equation}
    \label{response2}
    G(E)=\sum_n \frac{|R_n\rangle\langle L_n|}{\lambda_n-E} \, .
\end{equation}
For a Hermitian system we can choose normalized vectors $|R_n\rangle=|L_n\rangle$ such that $\|P_n\|=1$ with $P_n=|R_n\rangle\langle L_n|$. In this case the pole structure of $G(E)$ fully determines the response of the system. This is, however, no longer true for a non-normal Hamiltonian. In this case we can define normalized vectors $|\hat{R}_n\rangle=|R_n\rangle/\|R_n\|$ and $|\hat{L}_n\rangle=|L_n\rangle/\|L_n\|$ and choose the right and left eigenvectors to be bi-orthonormal $1=\langle L_n|R_n\rangle=\|R_n\| \|L_n\| \langle \hat L_n|\hat R_n\rangle$. For the projector this implies that 
\begin{equation}
    \label{response3}
    \|P_n\| = \frac{1}{|\langle \hat L_n|\hat R_n\rangle|} \, .
\end{equation}
But these normalized right and left eigenvectors can be nearly orthogonal for a non-normal matrix, $|\langle \hat L_n|\hat R_n\rangle|\ll 1$ in which case $\|P_n\|\gg 1$. I.e., for a non-normal matrix the pole structure given by the eigenspectrum does not fully determine the Green's function and thus the linear response. A large response is possible even far away from any pole if $\|P_n\|$ is large. We conclude that the eigenspectrum is the wrong object to infer the magnitude of the linear response of a non-normal system. 

Instead, we can use a singular value decomposition 
\begin{equation}
    \label{response4}
    G(E)=\sum_n\frac{|v_n\rangle\langle u_n|}{s_n(E)}
\end{equation}
where the vectors can now be chosen to be orthonormal so that $\||v_n\rangle\langle u_n|\|=1$. The norm (amplification) is therefore determined entirely by the singular values $s_n$. For a topologically protected singular value $s_1(E)$ we have, in particular
\begin{equation}
    \label{response5}
    \|G(E)\|_2 = \frac{1}{s_1(E)} \, .
\end{equation}
The response is determined by the smallest singular value $s_1(E)$ and $|u_1\rangle$ is the optimal source profile to drive this response while $|v_1\rangle$ is the profile of the response.}

To summarize, the eigenspectrum of non-normal operators such as the Hatano–Nelson Hamiltonian is highly unstable to generic perturbations and therefore cannot encode stable topological information. Although generalized Brillouin zone constructions can reproduce the open-boundary eigenspectrum, they necessarily miss the topological structure, which for finite systems is instead encoded in the singular-value spectrum. The non-Hermitian skin effect is a consequence of non-normality combined with non-reciprocal transport and is, in general, unrelated to topology. \js{In clean one-band models, such as the Hatano-Nelson model, this distinction is obscured because point-gap winding does predict the boundary reconstruction when cutting the clean system open. The skin modes occurring in the clean OBC case are, however, fragile to generic perturbations and therefore do not represent stable topological objects.} \js{This is not immediately obvious for the Hatano-Nelson model with OBC because this model is quasi-Hermitian}, which renders its spectrum unusually stable against certain local perturbations, masking the generic spectral instability underlying the skin effect. \js{Scalar one-band models such as the} Hatano–Nelson model therefore intertwine spectral instability and winding in a way that can easily lead to the mistaken impression that the non-Hermitian skin effect is a topological phenomenon. To separate them explicitly, we now turn to a Hatano--Nelson ladder, which is a \js{clean} two-band model where the winding structure and the non-normal pumping mechanism can be varied independently. This will allow us to provide examples of the NHSE without point-gap winding as well as of point-gap winding without the NHSE.

\section{Non-Hermitian skin effect without point-gap topology}
\label{Sec_ladder}
We present a simple model consisting of two coupled Hatano-Nelson chains, see Fig.~\ref{FigX}, which demonstrates that the NHSE \js{even in clean models} is, in general, unrelated to point-gap topology and is instead a consequence of non-normality and non-reciprocal couplings. \js{The goal is to present counterexamples that can be solved and understood analytically, not a classification of all types of clean multiband models where the scalar one-band boundary reconstruction theorem discussed in Sec.~\ref{II_D} breaks down. We note that in generic multiband models, if they can only be solved numerically, it is impossible to conclusively distinguish an extended state from a localized state with a localization length of the order or larger than the system size. This issue is particularly pronounced in non-Hermitian systems because multi-precision arithmetic is required to diagonalize a large ill-conditioned Hamiltonian matrix, severely limiting the system sizes that can be investigated.}
\begin{figure}
    \centering   
\begin{tikzpicture}[
    >=Latex,
    site/.style={circle, draw, fill=white, minimum size=7mm, inner sep=0pt},
    hopR/.style={->, thick},
    hopL/.style={<-, thick},
    interUp/.style={->, thick},
    interDown/.style={<-, thick},
    gammaHop/.style={<-, thick}
]

% spacing
\def\a{1.8}
\def\b{1.8}

% upper chain
\node[site] (A1) at (0,0) {$a_1$};
\node[site] (A2) at (\a,0) {$a_2$};
\node[site] (A3) at (2*\a,0) {$a_3$};
\node[site] (A4) at (3*\a,0) {$a_4$};

% lower chain
\node[site] (B1) at (0,-\b) {$b_1$};
\node[site] (B2) at (\a,-\b) {$b_2$};
\node[site] (B3) at (2*\a,-\b) {$b_3$};
\node[site] (B4) at (3*\a,-\b) {$b_4$};

% ellipses
\node at (3.5*\a,0) {$\cdots$};
\node at (3.5*\a,-\b) {$\cdots$};

% upper chain hopping
\draw[hopR] (A1) -- node[above=2pt] {$t_R$} (A2);
\draw[hopL] (A1) -- node[below=2pt] {$t_L$} (A2);

\draw[hopR] (A2) -- (A3);
\draw[hopL] (A2) -- (A3);

\draw[hopR] (A3) -- (A4);
\draw[hopL] (A3) -- (A4);

% lower chain hopping
\draw[hopR] (B1) -- node[above=2pt] {$\tilde t_R$} (B2);
\draw[hopL] (B1) -- node[below=2pt] {$\tilde t_L$} (B2);

\draw[hopR] (B2) -- (B3);
\draw[hopL] (B2) -- (B3);

\draw[hopR] (B3) -- (B4);
\draw[hopL] (B3) -- (B4);

% vertical inter-chain couplings
\draw[interUp]   (B1) -- node[left=2pt] {$t_U$} (A1);
%\draw[interDown] (B1) -- node[right=2pt] {$t_D$} (A1);

\draw[interUp]   (B2) -- (A2);
%\draw[interDown] (B2) -- (A2);

\draw[interUp]   (B3) -- (A3);
%\draw[interDown] (B3) -- (A3);

\draw[interUp]   (B4) -- (A4);
%\draw[interDown] (B4) -- (A4);

% diagonal gamma couplings: a_n -> b_{n+1}
%\draw[gammaHop] (B2) -- (A1);
%\draw[gammaHop] (B3) -- node[above right=-3pt] {$t_D\gamma$} (A2);
%\draw[gammaHop] (B4) -- (A3);

\end{tikzpicture}
\caption{Two Hatano-Nelson chains coupled by a non-reciprocal coupling $t_U$; see Eq.~\eqref{HN_doubled}.}
    \label{FigX}
\end{figure}
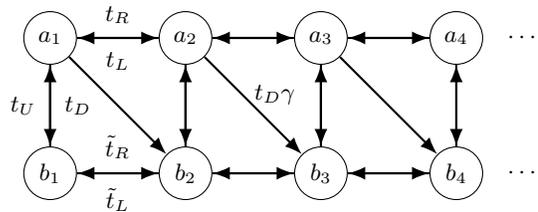
\begin{figure*}[t]
    \centering
    \includegraphics[width=0.325\textwidth]{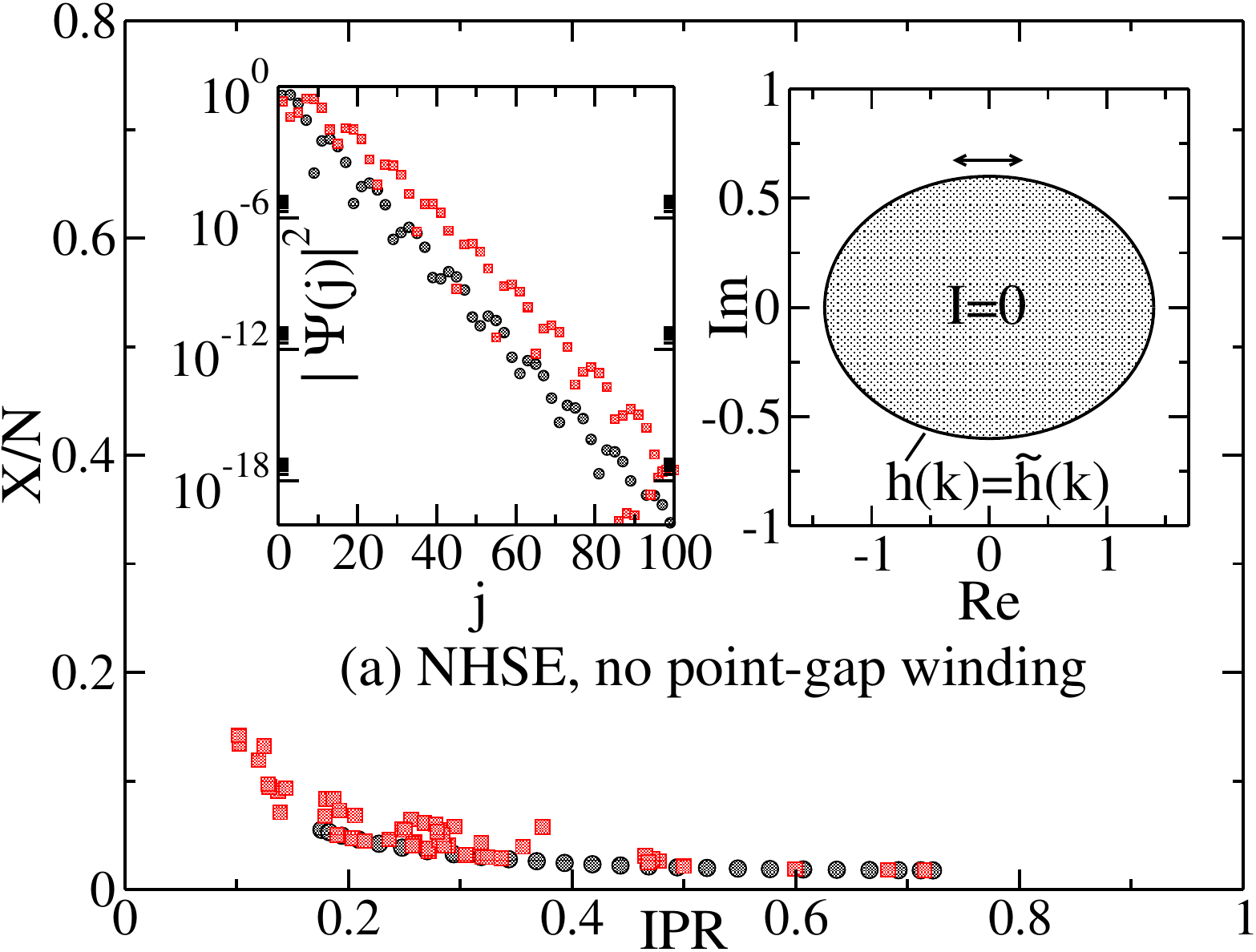}
    \includegraphics[width=0.325\textwidth]{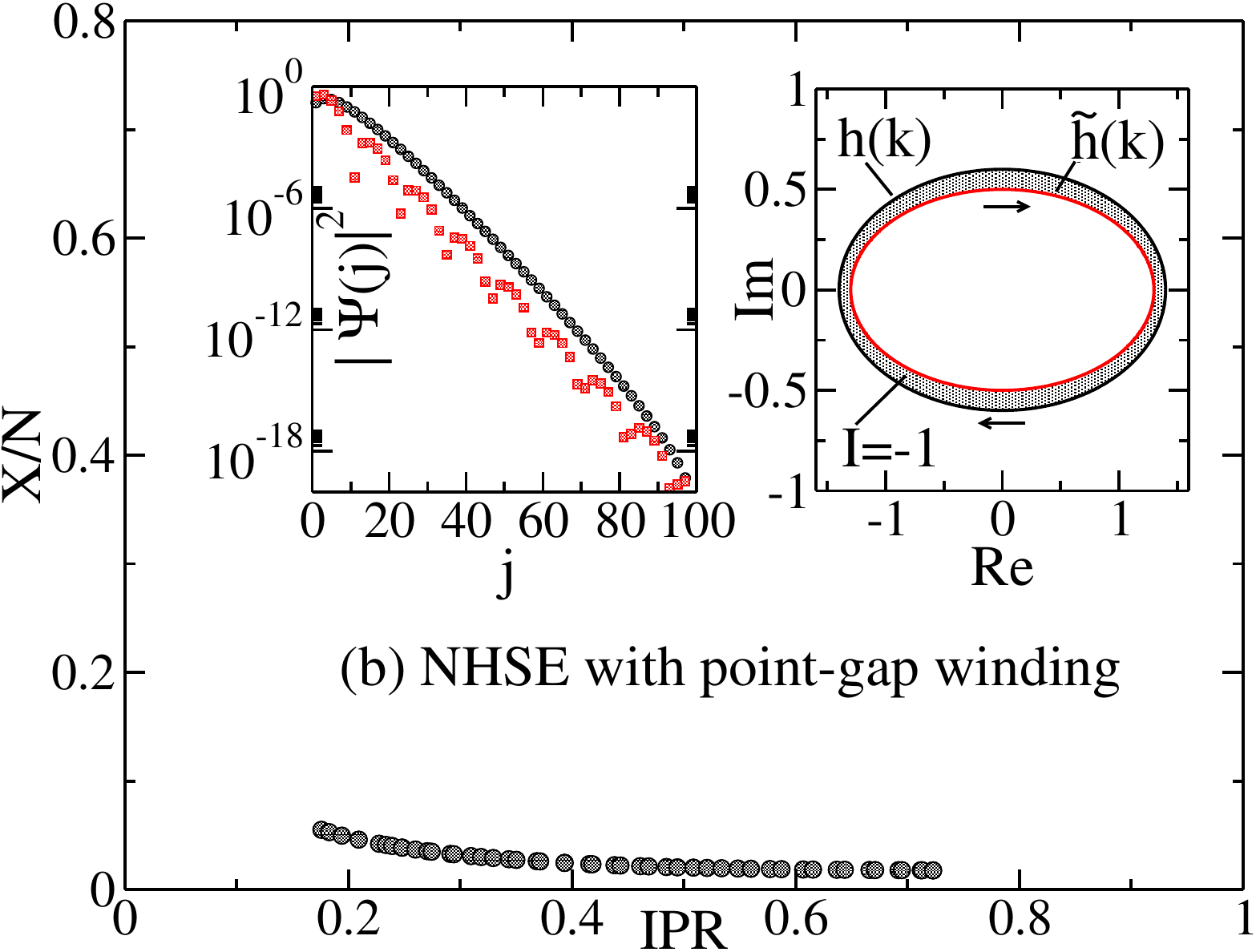}
    \includegraphics[width=0.325\textwidth]{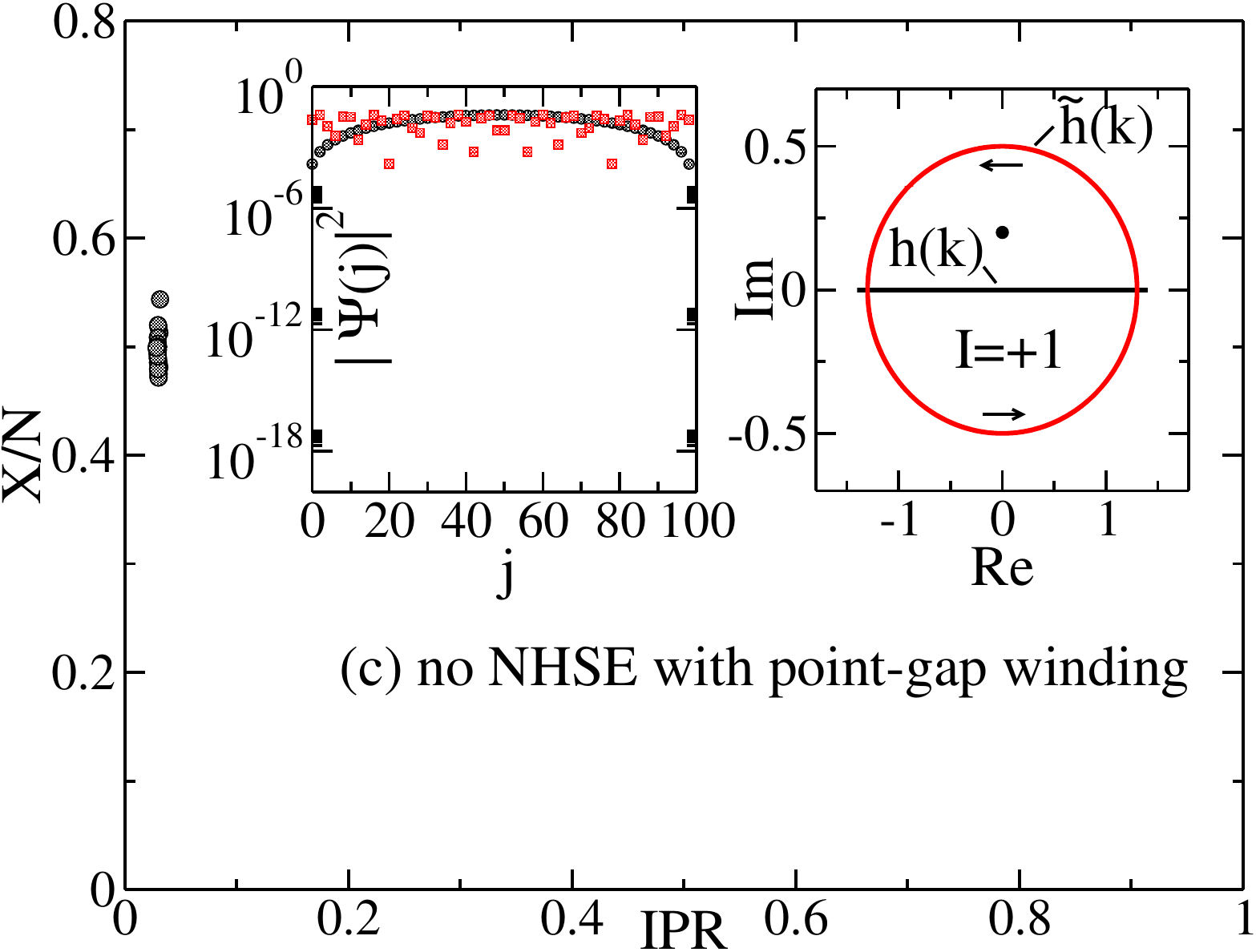}
    \caption{(a) $t_L=\tilde t_R=1$, $t_R=\tilde t_L=0.4$, $t_U=0.8$. Main panel: Eigenstates (black circles) and generalized eigenstates (red squares) are localized at the left boundary. Left inset: Spatial profiles of a representative eigenstate and generalized eigenstate. Right inset: Bloch Hamiltonian spectrum with winding $\mathcal{I}(E)=0$. (b) $\tilde t_R=0.9$ with all other parameters unchanged. The eigenstates remain localized at the left boundary. Right inset: Non-zero point-gap winding $\mathcal{I}(E)=-1$ around energies $E$ within the shaded area. (c) \js{$t_L=t_R=0.7$, $\tilde t_R=0.9$, $\tilde t_L=0.4$, $t_U=0.8$. Main panel: Eigenstates are now extended with centers clustered near $X/N\sim 0.5$. Left inset: Representative eigenstates show bulk character. Right inset: Non-zero point-gap winding persists. The dot marks $E=0.2\im$. Note that multiprecision arithmetic has been used.}}
    \label{FigX2}
\end{figure*}
The Hamiltonian \js{we will concentrate on} is given in second quantization by
\begin{eqnarray}
\label{HN_doubled}
H &=& \sum_{n}\Bigl\{ t_R\, a_{n+1}^\dagger a_n
+ t_L\, a_n^\dagger a_{n+1} + \tilde t_R\, b_{n+1}^\dagger b_n \Bigr. \\
&+& \Bigl. \tilde t_L\, b_n^\dagger b_{n+1} + t_U\, a_n^\dagger b_n 
%+ t_D\, b_n^\dagger (a_n+\gamma a_{n-1}) 
\Bigr\}  \nonumber
\end{eqnarray}
where $a_n^{(\dagger)}$ and $b_n^{(\dagger)}$ are the fermionic annihilation and creation operators on the two chains and $t_{L/R}$, $\tilde t_{L/R}$, and $t_{U}$ are real coupling parameters. If we Fourier transform this Hamiltonian, then we obtain the two-band Bloch Hamiltonian
\begin{equation}
    \label{HN_d2}
    H(k)=\begin{pmatrix}
        h(k)  & t_U \\
        %t_D(1+\gamma \e^{ik}) 
        0 & \tilde h(k)
    \end{pmatrix}
\end{equation}
with $h(k)=t_R\e^{ik}+t_L\e^{-ik}$ and $\tilde h(k)$ defined analogously. To calculate the winding number, the formula \eqref{winding} has to be generalized to
\begin{equation}
    \label{winding2}
    \mathcal{I}(E)=\frac{1}{2\pi \im}\int_0^{2\pi}dk\, \partial_k \ln(\det(H(k)-E))\, .
\end{equation}
If we Fourier transform the Bloch Hamiltonian using Eq.~\eqref{1dsymbol}, then we again obtain a Toeplitz operator of the form \eqref{Toplitz} but the entries are now $2\times 2$ matrices. We find, in particular,
\begin{equation}
    \label{HN_d3}
    h_0=\begin{pmatrix}
        0 & t_U \\
        %t_D & 0
        0 & 0
    \end{pmatrix},\;
    h_1=\begin{pmatrix}
        t_L & 0 \\
        0 & \tilde t_L
    \end{pmatrix},\;
    h_{-1}=\begin{pmatrix}
        t_R & 0 \\
        % t_D\gamma & \tilde t_R 
        0 & \tilde t_R 
    \end{pmatrix}
\end{equation}
with all other entries equal to zero. Here, the sites of the system are ordered as $(a_1,b_1,a_2,b_2,\cdots)$.

%\subsection{The triangular case}
\js{In the case considered here, where there is a hopping $t_U$ from the lower to the upper chain but not vice versa,} the real-space Hamiltonian matrix for OBC can be brought into upper triangular form, \js{allowing for an analytical solution.} If the sites are ordered as $(a_1,\cdots,a_n,b_1,\cdots,b_n)$, then the OBC Hamiltonian reads
\begin{equation}
    \label{HN_d4}
    H=\begin{pmatrix}
        H_{\rm HN} & t_U I \\
        0 & \tilde H_{\rm HN} 
    \end{pmatrix} \, .
\end{equation}
First, we note that the characteristic equation $\det(H-\lambda \mathbf{1})=\det(H_{\rm HN}-\lambda \mathbf{1})\det(\tilde H_{\rm HN}-\lambda \mathbf{1})=0$ shows that the eigenspectrum is given by $\sigma(H)=\sigma(H_{\rm HN})\cup \sigma(\tilde H_{\rm HN})$. The corresponding eigenvectors are given by $\Psi_j=(u_j\; 0)^T$ with $H_{\rm HN}u_j=E_ju_j$ and $\tilde\Psi_j=(\tilde u_j\; \tilde v_j)^T$ with $\tilde H_{\rm HN} \tilde v_j =\tilde E_j\tilde v_j$ and $\tilde u_j = -(H_{\rm HN}-\tilde E_j\mathbf{1})^{-1}t_U\tilde v_j$ if $\sigma(H_{\rm HN})\cap \sigma(\tilde H_{\rm HN})=\emptyset$. The upper-triangular coupling $t_U$ thus does not change the spectra of the two Hatano-Nelson chains but induces a one-way hybridization of the eigenstates of the $\tilde H_{\rm HN}$ chain into the $H_{\rm HN}$ chain. \js{If} $\tilde E_j$ approaches the spectrum of $H_{\rm HN}$ with increasing system size, the norm of $(H_{\rm HN}-\tilde E_j\mathbf{1})^{-1}$ becomes large and the upper component of the eigenstate $\tilde\Psi_j$ will dominate. \js{In this case}, the eigenstate \js{for $t_R\neq t_L$} will therefore become localized at the same edge as the eigenvector $\Psi_j$. As can be seen from Fig.~\ref{FigX}, the system acts like a pump, pushing all the weight of the eigenstates to the edge to which the dominant hopping process in the upper chain is pointing.

Of particular interest is the case when the non-reciprocities in the two chains are equal but opposite, $t_R=\tilde t_L$ and $t_L=\tilde t_R$. In this case, the spectra of $H_{\rm HN}$ and $\tilde H_{\rm HN}$ are the same and each eigenvalue of $H$ is twofold degenerate. However, only one eigenvector $\Psi_j=(u_j\; 0)^T$ with $H_{\rm HN}u_j=E_ju_j$ exists for each eigenvalue $E_j$. I.e., for the $2N$ eigenvalues in total there are only $N$ eigenvectors. This means that the matrix $H$ is defective and can only be brought into Jordan normal form, not into diagonal form. To achieve Jordan normal form, we can extend the $N$ eigenvectors $\Psi_j$ to a complete basis by supplementing each eigenvector by a generalized eigenvector $\Psi'_j$ which fulfills the equation $(H-E_j\mathbf{1})\Psi'_j=\Psi_j$. Because the eigenvectors $\Psi_j$ are exponentially localized at the boundary, this inhomogeneous equation produces generalized eigenvectors $\Psi_j'$ with the same boundary-localized profile (up to polynomial prefactors). This is shown in Fig.~\ref{FigX2}(a), which demonstrates that the center of all the eigenstates and generalized eigenstates are localized at the left boundary. This means that the NHSE is present. However, due to the equal but opposite non-reciprocities along the two chains, the system has no point-gap winding {\it around any energy $E$ in the complex plane}. From Eq.~\eqref{winding2}, we see that $\mathcal{I}(E) =\text{wind}(h(k)-E) + \text{wind}(\tilde h(k)-E)=0$ because $\tilde h(k)=h(-k)$. The windings are equal and opposite to each other, see the inset of Fig.~\ref{FigX2}(a), and therefore cancel. This constitutes an example where the NHSE is present but there is no point-gap winding. 

The case of exactly equal and opposite non-reciprocal couplings is fine-tuned, leading to a defective matrix. A natural question to ask is, therefore, what happens if we keep the triangular structure of $H$ but move away from this special point. In this case the matrix is no longer defective, and the eigensystem of $H$, given by the vectors $\Psi_j$ and $\tilde \Psi_j$ $(j=1,\cdots,N)$, forms a complete basis. As shown in Fig.~\ref{FigX2}(b), this eigensystem remains boundary localized but now there is a point-gap winding $\mathcal{I}(E)=-1$ present around certain energies $E$ in the complex plane, see the shaded area in the right inset. If the NHSE is of topological origin, then one would expect that changing from a fine-tuned case without point-gap winding to one with point-gap topology would drastically alter the localization properties of the eigenstates. That this is not the case shows that the origin of the NHSE is instead the strong non-normality of the triangular matrix \eqref{HN_d4} together with the non-reciprocal couplings. The special point with no point-gap winding is fully representative of this class.

\js{Finally, we consider a case where energies $E$ exist with $\mathcal{I}(E)\neq 0$ but there is no skin effect. We choose the upper chain to have reciprocal couplings $t_R=t_L$ while the lower chain has non-reciprocal couplings $\tilde t_R\neq \tilde t_L$. The OBC eigenenergies are then $E_j=2\sqrt{t_R t_L}\cos q_j$ and similarly for $\tilde E_j$. We choose parameters such that the OBC spectrum $\sigma(\tilde H_{\rm HN})$ of the lower non-reciprocal chain lies inside the spectrum $\sigma(H_{\rm HN})$ of the upper reciprocal chain. The eigenvectors $\Psi_j=(u_j \; 0)^T$ with $H_{\rm HN} u_j = E_j u_j$ are eigenvectors of the reciprocal chain and therefore extended. The question is then whether the eigenstates $\tilde\Psi_j=(\tilde u_j\; \tilde v_j)^T$ with $\tilde H_{\rm HN} \tilde v_j = \tilde E_j \tilde v_j$ and $\tilde u_j = -(H_{\rm HN}-\tilde E_j\mathbf{1})^{-1}t_U\tilde v_j$ for $\tilde E_j\notin \sigma(H_{\rm HN})$ are localized or extended. The component $\tilde v_j$ is an eigenstate of the non-reciprocal chain $\tilde H_{\rm HN}$ and therefore skin-localized. However, if the spectrum $\sigma(\tilde H_{\rm HN})$ lies inside the spectrum $\sigma(H_{\rm HN})$, then the resolvent $(H_{\rm HN}-\tilde E_j\mathbf{1})^{-1}$ is large: since $\tilde v_j$ is exponentially localized while the
eigenmodes of $H_{\rm HN}$ are extended, the overlaps are $\mathcal{O}(N^{-1/2})$ and the nearest level lies at a distance $\mathcal{O}(1/N)$, so that $\|\tilde u_j\|\sim t_U\sqrt{N}\,
\|\tilde v_j\|$. The component $\tilde u_j$ is therefore extended and dominant, with the skin-localized admixture carrying a relative weight $\sim 1/(t_U^2 N)$ and the inverse
participation ratio scaling as $1/N$. All eigenstates of the system in this regime are therefore extended despite the fact that energies with a non-zero point-gap winding exist, see Fig.~\ref{FigX2}(c). 

This example shows one way how the scalar one-band boundary reconstruction theorem can fail in the clean multiband case: the winding based on $\det(H(k)-E)$ still counts roots but this root count no longer controls the entire wave function. The skin-localized component of the wave function can be sub-dominant. Physically, the NHSE disappears because the skin modes hybridize with a reciprocal channel.}

\begin{figure}[t]
    \centering
    \includegraphics[width=0.99\columnwidth]{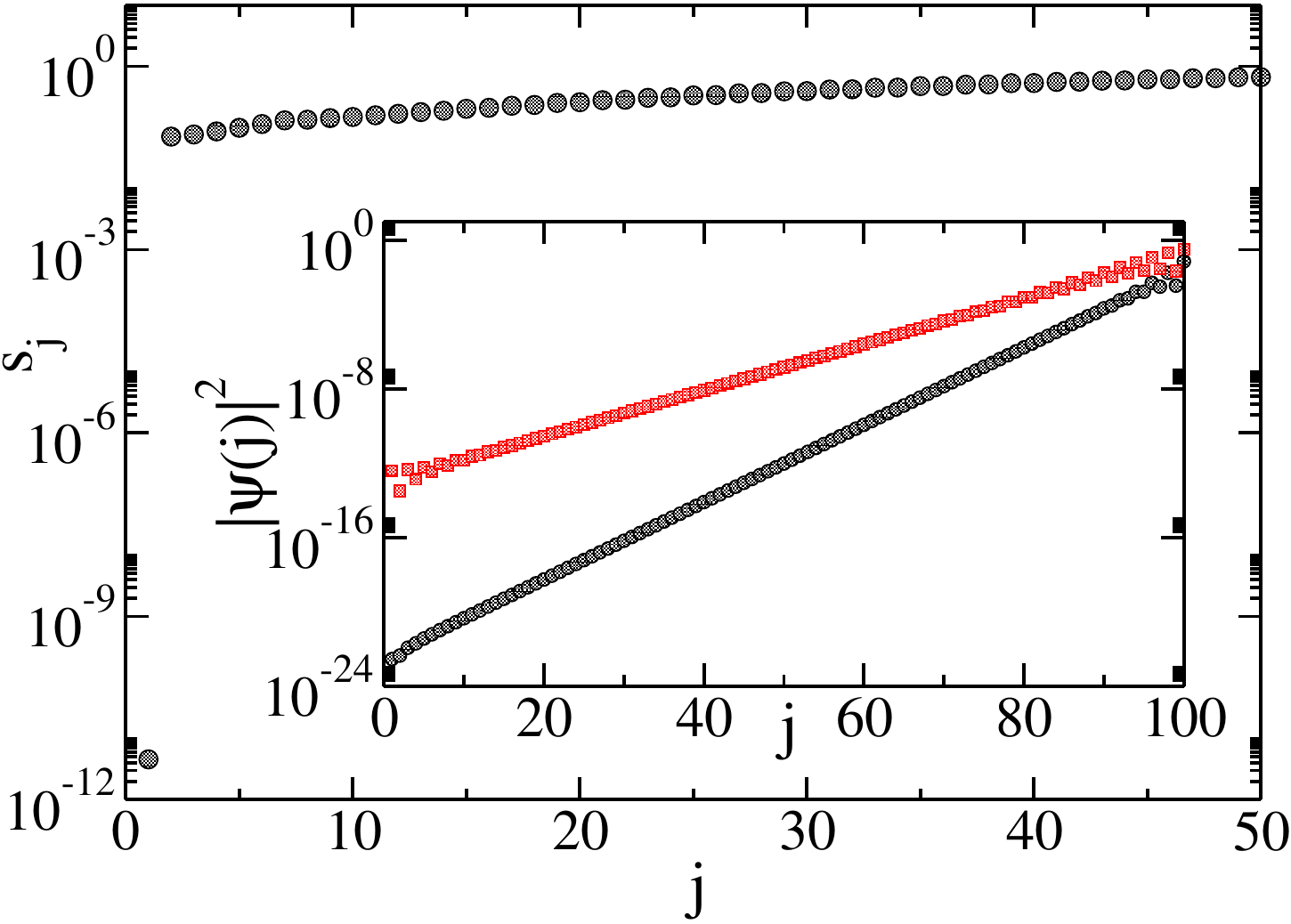}
    \caption{Singular value spectrum of $H-E$ for $E=0.2\im$, see black dot in Fig.~\ref{FigX2} right inset. There is one topologically protected singular value. The exponentially localized components of the corresponding singular vector along the two chains are shown in the inset.}
    \label{FigX3}
\end{figure}

However, since there is winding around certain energies in the complex plane, there are topologically protected singular vectors, which indicate the existence of protected boundary-localized eigenstates in the thermodynamic limit. As an example, we show in Fig.~\ref{FigX3} the singular value spectrum of $H-E$ with $E=0.2\im$ (black dot in the right inset of Fig.~\ref{FigX2}(c)).
Consistent with the winding $\mathcal{I}(E)=+1$, there is a singular value that is separated from the bulk by a gap and that tends to zero in the thermodynamic limit. The corresponding singular vector shows exponentially localized components on both of the coupled Hatano-Nelson chains. \js{Experimentally, this boundary localized metastable state at $E=0.2\im$ will be visible in linear response, see Eq.~\eqref{response4}, despite the fact that the finite system has no boundary localized eigenmodes.}

To summarize, these examples show that \js{even in the clean case} the NHSE and point-gap topology are two properties of non-Hermitian systems which are, in general, independent of each other. The NHSE requires a non-normal matrix, which makes the spectrum sensitive to a change in boundary conditions, and non-reciprocal couplings, which lead to the eigenstates preferentially accumulating on one of the edges. 

Finally, we note that it has been suggested that in systems with subspace symmetry, the winding number of the block of the Bloch Hamiltonian representing this subspace is the relevant quantity \cite{ShimomuraTakami}. For the triangular matrix \eqref{HN_d4} these would be the winding numbers associated with the diagonal blocks. However, this interpretation is not robust. \js{First, in the example shown in Fig.~\ref{FigX2}(c) one of the blocks does have a non-zero winding number yet all states are extended. Second,} the NHSE persists under clean perturbations that break the triangular (or subspace) structure, indicating that such a symmetry is not required for boundary localization. \js{Third}, in the strictly triangular case, the eigenstates are controlled by the underlying Hatano–Nelson blocks, and are therefore unstable to generic random disorder, which destroys the NHSE. These observations show that subspace-based interpretations do not provide a stable or generic explanation of the NHSE.

\section{Conclusions}
\label{Sec_Concl}
In this work, we have revisited the relation between the non-Hermitian skin effect (NHSE) and point-gap topology from the perspective of spectral stability. \js{We have stressed that it is important to clearly distinguish between the question of boundary reconstruction when cutting a clean system open and the question of what the stable topological properties of a non-Hermitian system are.} Using the Hatano-Nelson model as a paradigmatic example, we have illustrated the general property that the eigenspectrum of a non-normal Hamiltonian is highly sensitive to boundary conditions and generic perturbations, and therefore does not constitute a stable object capable of encoding topological information. Instead, the topological properties are captured by the index of the corresponding Toeplitz operator, which indicates exact boundary eigenmodes for the semi-infinite system. For a finite system, the existence of such topological modes is reflected in the singular-value spectrum, which remains stable under small generic perturbations. \js{Experimentally, the topologically protected singular values lead to a strong amplification in linear response, with the source and response profiles determined by the corresponding singular vectors.}

To further clarify the \js{relation between point-gap winding and boundary reconstruction in clean non-Hermitian systems---which is distinct from the question of stable topological protection summarized above---}we constructed a two-band extension of the Hatano-Nelson model. \js{This model shows that the equivalence between the existence of an energy with non-zero point-gap winding and the existence of a macroscopic number of skin states in general only holds in clean scalar one-band models. In particular, the clean two-band Hatano-Nelson model} allowed us to identify three distinct regimes: (i) the macroscopic accumulation of localized states at an edge in the absence of point-gap winding, (ii) boundary-localized states in the presence of point-gap winding, and (iii) no boundary-localized states despite nonzero point-gap winding. Taken together, these examples demonstrate that \js{even in clean non-Hermitian systems} the NHSE and point-gap topology are, in general, independent properties.

Our results clarify that the NHSE originates from the non-normality of the Hamiltonian and the associated spectral instability, rather than from topological winding. The commonly observed correspondence between spectral winding and boundary localization relies implicitly on translational invariance and an associated generalized Bloch description \js{and generically only holds in clean scalar one-band models}. Since translational invariance is not a topological property, this correspondence, \js{even in the scalar one-band case,} is not generic: once perturbations are introduced, the eigenspectrum of non-normal operators becomes unstable and no longer reflects the winding of the (generalized) Bloch Hamiltonian.

More broadly, our findings highlight the importance of distinguishing between stable and unstable spectral quantities in non-Hermitian systems. While point-gap topology remains well-defined at the level of the Bloch Hamiltonian and is reflected in the singular-value spectrum, it is, in general, unrelated to the behavior of the eigensystem of finite non-Hermitian matrices. We therefore conclude that the NHSE is not a topological phenomenon, and that a consistent formulation of a bulk-boundary correspondence in non-Hermitian systems must be based on Toeplitz operator theory and the singular-value spectrum, which are stable objects, rather than the eigenspectrum.

\begin{acknowledgments}
The author acknowledges support by the National Science and Engineering Research Council (NSERC) of Canada through the Discovery Grants program.
\end{acknowledgments}

%\clearpage
%\bibliography{Literatur}
%

\end{document}